\documentclass[a4paper,11pt]{article}
\usepackage{pos}
\usepackage[export]{adjustbox}
\usepackage{mciteplus}
\title{Recent results on (semi)-leptonic $D$ decays and charm baryons at BESIII}
\usepackage{ifthen}
\newboolean{articletitles}
\setboolean{articletitles}{true}

\author*[a,b]{Alex Gilman}

\onbehalf{for the BESIII collaboration}

\affiliation[a]{Department of Physics, University of Cincinnati,\\
  Cincinnati, Ohio 45221, USA}

\affiliation[b]{Department of Physics, University of Oxford,\\
Oxford OX1 3PU, UK\\
Presented at the 32nd International Symposium on Lepton Photon Interactions at High Energies, Madison, Wisconsin}

\emailAdd{alexander.leon.gilman@cern.ch}

\usepackage{xspace} 
\def\beq{\begin{equation}}
\def\eeq{\end{equation}}

\def\psipp{\psi(3770)}

\def\DSTO{\ensuremath{D^{*0}}\xspace}

\def\aDSTO{\offsetoverline{D}^{*0}}

\def\ee{e^+e^-}

\def\Ecm{E_{\text{cm}}}

\def\invfb{\text{ fb}^{-1}}

\def\GeV{\text{ GeV}}

\def\thebaroffset{0.1em}
\newcommand{\offsetoverline}[2][\thebaroffset]{\kern #1\overline{\kern -#1 #2}}%

\def\D       {{\ensuremath{D}}\xspace}

\def\Dp      {{\ensuremath{\D^+}}\xspace}
\def\Dm      {{\ensuremath{\D^-}}\xspace}

\def\DpDm    {\ensuremath{\Dp {\kern -0.16em \Dm}}\xspace}

\def\Dstarz  {\DSTO\xspace}
\def\Dstarzb {\aDSTO\xspace}

\def\Ppsi        {\ensuremath{\psi}\xspace}  
\def\psipp  {{\ensuremath{\Ppsi(3770)}}\xspace}

\def\DstarzDstarzb    {\ensuremath{\Dstarz {\kern -0.07em \Dstarzb}}\xspace}

\newcommand{\aunit}[1]{\ensuremath{\text{\,#1}}}       
    
\def\fb   {\ensuremath{\aunit{fb}}\xspace}
\def\invfb   {\ensuremath{\fb^{-1}}\xspace}
 
\abstract{The BESIII collaboration has collected the world's largest datasets at the energy thresholds for producing a variety of open-charm hadrons.  Most recently, the BESIII collaboration has collected datasets that significantly increase the number of $D^0$, $D^+$, and $\Lambda_c^+$ hadrons for analysis, further improving on previous datasets collected by BESIII. These proceedings highlight recent results that leverage these datasets to study the pure leptonic and semileptonic decays of these hadrons, as well as study the polarization of baryon pairs produced in electron-positron collisions. }

\FullConference{%
  Lepton-Photon 2025\\
  Aug. 25- Aug. 29 2025\\
  Madison, Wisconsin, USA
}

\begin{document}
\maketitle

%\section{Introduction}

%\section{The BEPCII Collider and BESIII Detector}

The BESIII experiment~\cite{BESIIIDetector} records data of symmetric $e^+e^-$ collisions provided by the Beijing Electron Positron Collider Mk. II \cite{BEPCII} (BEPCII). BEPCII produces collisions at center-of-mass energies of $2-5 \text{ GeV}$ with a design luminosity of $10^{33}$ $\text{cm}^{-2}\text{s}^{-1}$ achieved in April 2016. The BESIII detector covers $93\%$ of the full solid angle, and is equipped with gaseous tracking system, a plastic scintillator time-of-flight system for particle identification, a caesium-iodide calorimeter, and a resistive-plate-chamber muon system.
BEPCII and BESIII finished construction in 2008, upgrading on the original Beijing Electron-Positrion Collider and BESII detector. BESIII began taking data in 2009, and has collected many large datasets in this energy regime\cite{BESIII:2020nme}. These proceedings report results on $D^0$ and $D^+$ decays\footnote{Charge conjugation is implied throughout these proceedings, except where explicitly stated. } $20.3\;\invfb$ sample collected at the \psipp threshold, whose collection was completed in 2024. This contains the previous largest sample at this energy, a sample of $2.9\invfb$\;collected in 2011. These proceedings also report results which use a $6.4\invfb$ BESIII sample collected between 2022-2024 at center-of-mass energies between $\Ecm=4.60-4.95\GeV$, which contain sizable samples of $\ee\to\Lambda_c^+\Lambda_c^-$.

Electron-positron collisions near open-charm pair-production energy thresholds allow for various unique studies of fundamental physics. These proceedings summarize recent results on two classes of such measurements: pure leptonic and semileptonic decays of charmed hadrons, and other results relating to charmed baryons. Precision measurements of the leptonic and semileptonic decays of charmed hadrons enable:tests of CKM unitarity through precision measurements of $|V_{cd}|$ and $|V_{cs}|$; studies of non-perturbative QCD through determinations of decay constants and form factors that can be compared with lattice QCD and other theoretical calculations; tests of lepton flavor universality by comparing decay rates to electrons versus muons; and a laboratory for light hadron physics, including the structure of scalar mesons such as the $a_0(980)$.

The double-tag method\cite{MARK-III:1987jsm,ARGUS:1990hfq} exploits the pair production of charmed hadrons to measure the branching fractions of decay modes with particles that are difficult to reconstruct, such as neutrinos. First, one charmed hadron (the ``tag'') is reconstructed through a clean hadronic decay mode. The presence of the tag then guarantees the existence of another charmed hadron of opposite flavour in the event, whose decay can be studied inclusively by searching for signal particles in the remainder of the event. Missing particles such as neutrinos can be identified through kinematic variables such as the missing mass squared $M^2_{\rm miss} = (p_{e^+e^-} - p_{\rm tag} - p_{\rm visible})^2$ or the missing energy-momentum $U_{\rm miss} \equiv E_{\rm miss} - p_{\rm miss}$, where $p_{e^+e^-}$ is the four-momentum of the initial $e^+e^-$ system. The branching fraction of the signal decay is then given by
\begin{equation}
\mathcal{B}(D \to {\rm signal}) = \frac{N_{\rm signal}/\epsilon_{\rm tag~\&~signal}}{N_{\rm tag}/\epsilon_{\rm tag}},
\end{equation}
where $N_{\rm tag}$ and $N_{\rm signal}$ are the yields of tag and signal decays, respectively, and $\epsilon$ denotes the corresponding efficiencies. This method offers several advantages compared to only reconstructing the signal decay channel: combinatorial backgrounds are greatly reduced, and the full kinematic information of the recoiling system is accessible for differential measurements.

\begin{figure}[hbtp]
    \centering
    \begin{tabular}{cc}
    \includegraphics[width=0.45\linewidth, valign=c]{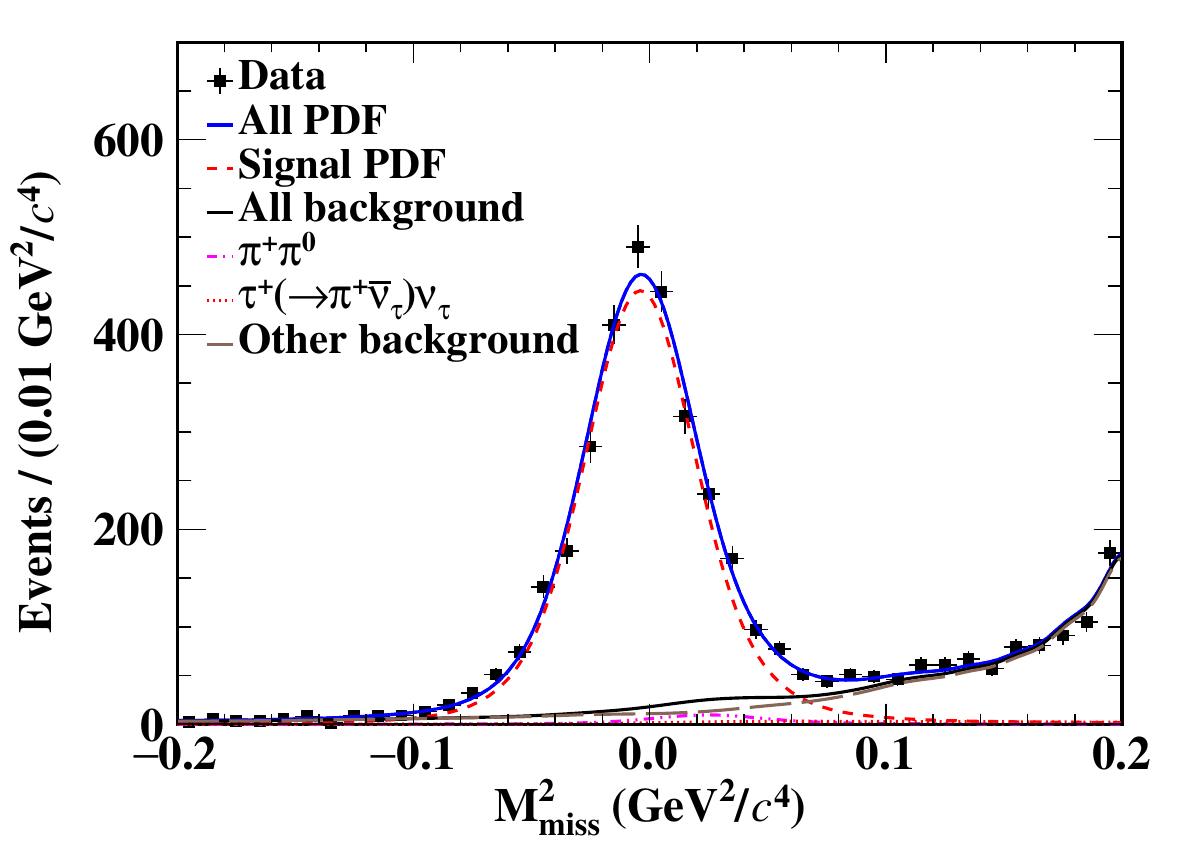}
         &    \includegraphics[width=0.5\linewidth, valign=c]{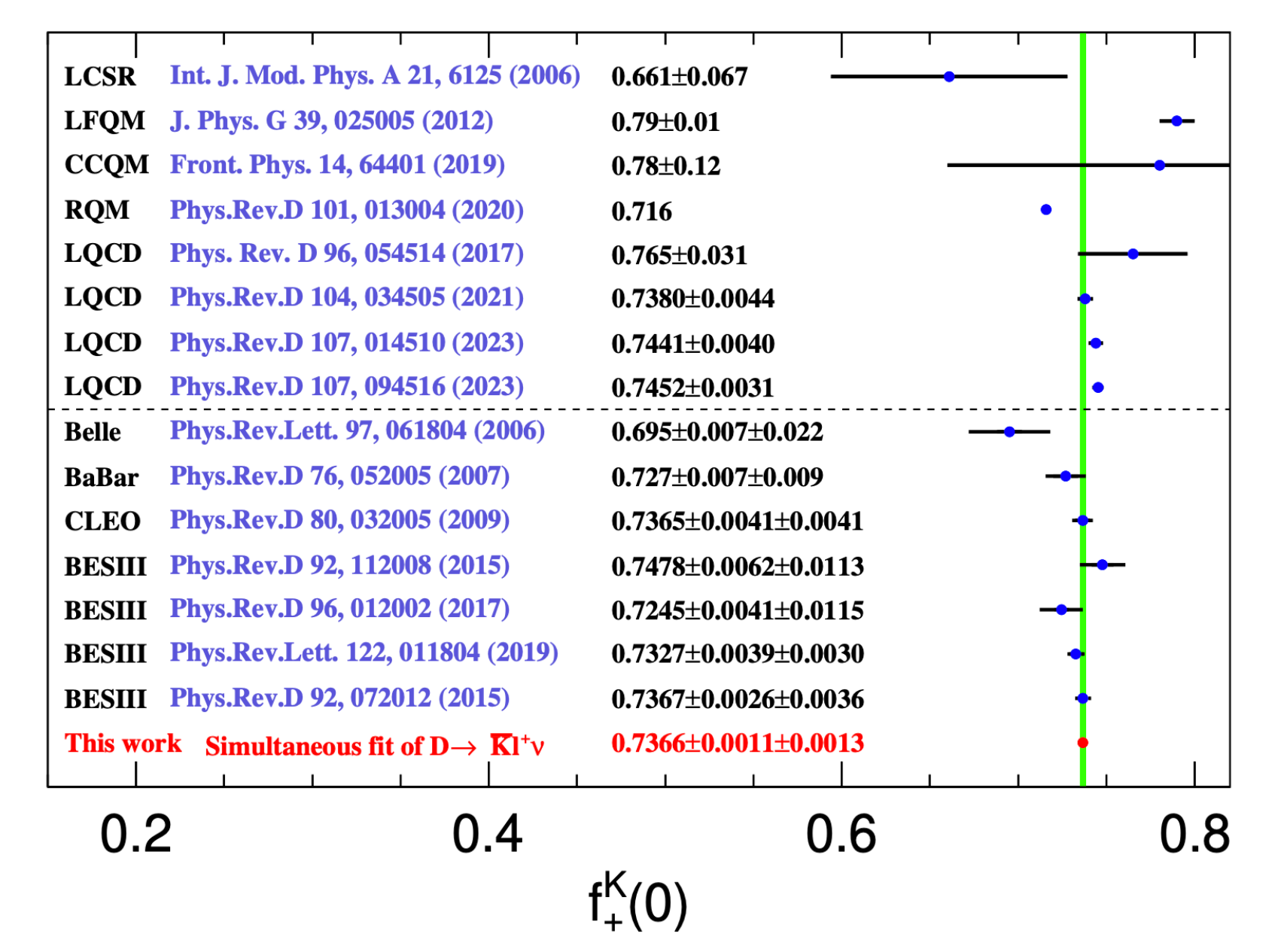}
    \end{tabular}

    \caption{(Left) Fit to determine $D^+\to \mu^+\nu$ signal yields from Ref.~\cite{gb8v-4rnh}. (Right) Comparison of $f^K_+(0)$ determined in Ref.~\cite{BESIII:2024slx} to various other experimental and theoretical determinations. }
    \label{fig:firstfig}
\end{figure}

The full 20.3~fb$^{-1}$ dataset collected at $E_{\rm CM} = 3.773$~GeV was analyzed to measure the purely leptonic decay $D^+ \to \mu^+ \nu_\mu$~\cite{gb8v-4rnh}. Eight $D^-$ tag modes were employed: $K^+\pi^-\pi^-$, $K^0_S\pi^-$, $K^+\pi^-\pi^-\pi^0$, $K^0_S\pi^-\pi^0$, $K^0_S\pi^-\pi^+\pi^-$, $K^+K^-\pi^-$, $\pi^+\pi^-\pi^-$, and $K^+\pi^-\pi^-\pi^-\pi^+$. Tag yields were determined from fits to the beam-constrained mass $M_{\rm BC} \equiv \sqrt{E^2_{\rm beam} - |\vec{p}_D|^2}$. Muon candidates were identified through selections on electromagnetic calorimeter energy deposition and muon chamber penetration depth. Signal yields were extracted from a fit to the $M^2_{\rm miss}$ distribution, with the dominant peaking backgrounds arising from $D^+ \to \pi^0\mu^+\nu$ and $D^+ \to \tau^+\nu \to \pi^+\pi^0\nu\nu$ decays. The final sample, and the fit to determine the signal yields is shown in Fig.~\ref{fig:firstfig}.  The measured branching fraction is $\mathcal{B}(D^+ \to \mu^+\nu_\mu) = (3.98 \pm 0.08 \pm 0.04) \times 10^{-4}$, representing a 2.3-fold improvement in precision over previous measurements. Using the $D^+$ lifetime from PDG2022\cite{ParticleDataGroup:2020ssz}, this results in the determination $f_{D^+}|V_{cd}| = (47.53 \pm 0.48 \pm 0.24 \pm 0.12_{\rm ext})$~MeV. This result is complemented by a new measurement of $D^+ \to \tau^+\nu$ using 7.9~fb$^{-1}$ of data at the $\psi(3770)$ resonance~\cite{BESIII:2024vlt}, which comprises a subset of the full 20.3~fb$^{-1}$ dataset.

The golden channels for $|V_{cs}|$ determination, $D^0 \to K^-(e^+/\mu^+)\nu$ and $D^+ \to K^0_S(e^+/\mu^+)\nu$, were analyzed using a 7.9~fb$^{-1}$ sample collected at $E_{\rm CM} = 3.773$~GeV~\cite{BESIII:2024slx}. Six $\bar{D}^0$ tag modes and six $D^-$ tag modes were employed. Signal yields were extracted from fits to $U_{\rm miss}$, with peaking backgrounds in the muon sample arising from challenges in low-momentum $\mu/\pi$ separation. The measured branching fractions achieve relative precisions of 0.5\%--1.1\%, enabling the most precise test of lepton flavor universality in charm decays: $R^{D^+}_{\mu/e} = 0.978(7)_{\rm stat}(13)_{\rm syst}$ and $R^{D^0}_{\mu/e} = 0.971(4)_{\rm stat}(6)_{\rm syst}$, consistent with the Standard Model expectation of $0.975(1)$~\cite{Riggio:2017zwh}.The differential decay rate ${\rm d}\Gamma/{\rm d}q^2$, where $q^2$ is the invariant mass squared of the lepton-neutrino system, was measured and the form factors were parametrized using a power-series expansion. A simultaneous fit to both $D^0$ and $D^+$ modes yields $f^{K}_+(0)|V_{cs}| = 0.7171(11)_{\rm stat}(13)_{\rm syst}$. Taking $|V_{cs}|$ from the PDG2022 global fit \cite{ParticleDataGroup:2020ssz}, the form factor $f^K_+(0) = 0.7366 \pm 0.0011 \pm 0.0013$ is obtained. This result is compared to previous experimental and theoretical determinations in Fig.~\ref{fig:firstfig}, where a $\sim2\sigma$ tension is observed with the most recent lattice results\cite{Parrott:2022rgu,FermilabLattice:2022gku}.

The semileptonic decays $D^+ \to K^0_S\pi^0\ell^+\nu$, where $\ell^+=e^+\text{ or } \mu^+$, were studied using the full 20.3~fb$^{-1}$ dataset at the $\psipp$ resonance with six $D^-$ tag modes~\cite{w9vz-4fq9}. Signal yields were determined from fits to $U_{\rm miss}$ for both $e^+$ and $\mu^+$ final states. In these decays, the $K\pi$ system is seen to be dominated by the $K^*(892)$ resonance, whose amplitude structure and form factor ratios were extracted from a five-dimensional fit to the variables $m^2_{K\pi}$, $q^2$, $\cos\theta_\ell$, $\cos\theta_K$, and $\chi$, where $\theta_\ell$ and $\theta_K$ are helicity angles and $\chi$ is the angle between decay planes. A smaller contribution from $S$-wave $K\pi$ decays is also observed.  The fit yields the $K^*(892)$ mass and width, the form factor ratios $r_2 = A_1(0)/A_2(0)$ and $r_V = V(0)/A_1(0)$, the shape parameters of the form factors, and the $K\pi$ $S$-wave fraction $f_{S\text{-wave}}$. Observables associated with CP violation, lepton flavor universality, and forward-backward asymmetries  were measured in different $q^2$ regions of the sample. All CP asymmetries and angular observables are consistent Standard Model predictions, and the ratio $R_{\mu/e}$ is consistent with unity across all $q^2$ bins. This analysis is part of a recent series of BESIII measurements of $D \to KX\ell\nu$ decays, including $D^0 \to K^0_S\pi^-\mu^+\nu$~\cite{zfxr-dlzg}, $D^+ \to \bar{K}_1(1270)^0\ell^+\nu$~\cite{BESIII:2025yot}, and $D^0 \to K^0_S\pi^-e^+\nu$~\cite{BESIII:2024xjf}.

A new measurement of the semileptonic decay $D^0 \to a_0(980)^-e^+\nu$ was performed using a 7.9~fb$^{-1}$ sample at the $\psipp$ resonance with six $D^-$ tag modes~\cite{BESIII:2025ujq}, providing the first analysis of the differential decay rate and the first experimental determination of the $D\to a_0$ form factor. The $a_0(980)^-$ state was studied through its decay to $\eta\pi^-$, with the $\eta$ reconstructed via $\gamma\gamma$. Signal yields were extracted from a two-dimensional fit to $U_{\rm miss}$ and $m_{\eta\pi^-}$. The $a_0(980)$ resonance was parametrized using the Flatt\'e formalism, and the contribution from non-resonant $\eta\pi^-$ production was found to be negligible. The measured branching fraction is $\mathcal{B}(D^0 \to a_0(980)^-[{\to}\eta\pi^-]e^+\nu_e) = (0.86 \pm 0.17 \pm 0.05) \times 10^{-4}$. The differential decay rate ${\rm d}\Gamma/{\rm d}q^2$ and the form factor $f^{a_0}_+(q^2)$ were measured as functions of $q^2$. Using $|V_{cd}|$ from PDG2024\cite{ParticleDataGroup:2024cfk}, the form factor was determined to be $f^{D\to a_0}_0 = (0.550 \pm 0.056 \pm 0.013)$. This result provides important discriminating power between different theoretical models of the $a_0(980)$ structure\cite{Soni:2020sgn,Momeni:2022gqb,Cheng:2017fkw,Huang:2021owr,Wu:2022qqx}.

\begin{figure}[hbtp]
    \centering
    \begin{tabular}{cc}
    \includegraphics[width=0.55\linewidth, valign=c]{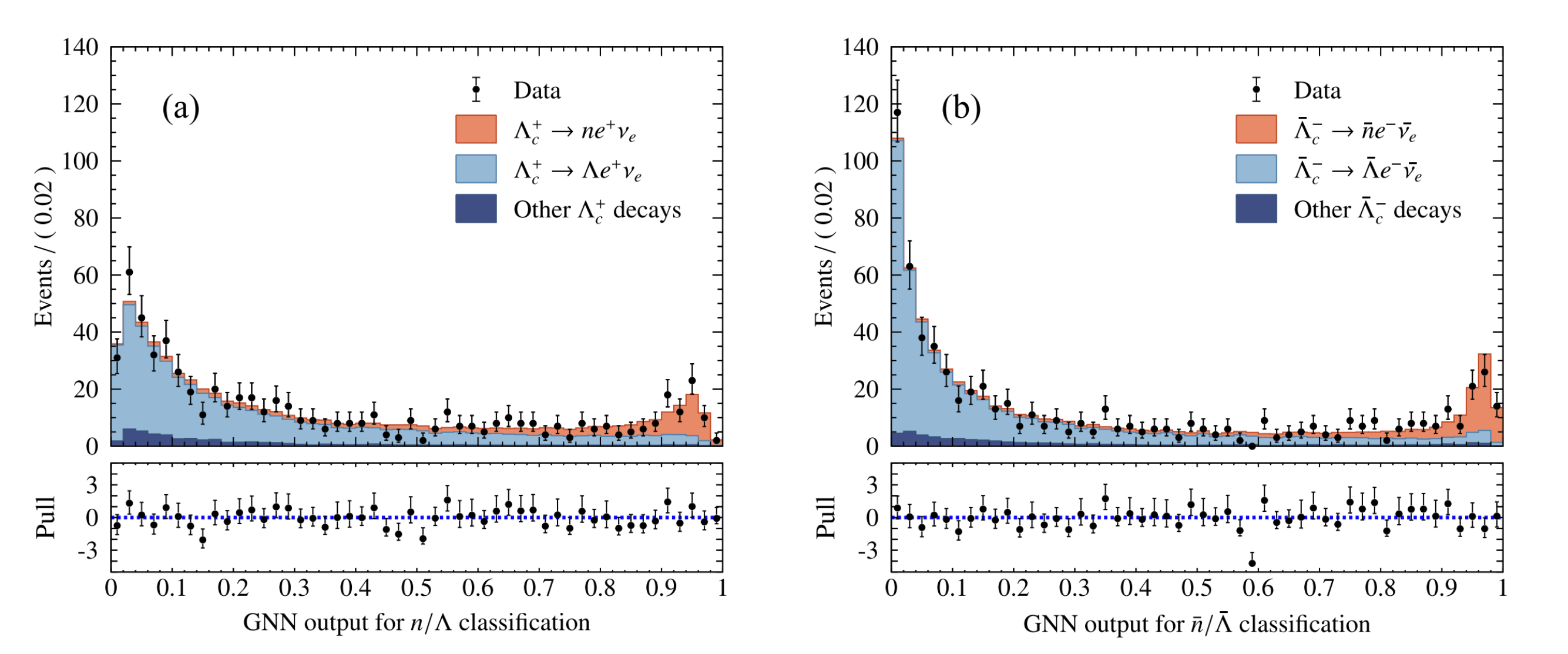}
         &    \includegraphics[width=0.35\linewidth, valign=c]{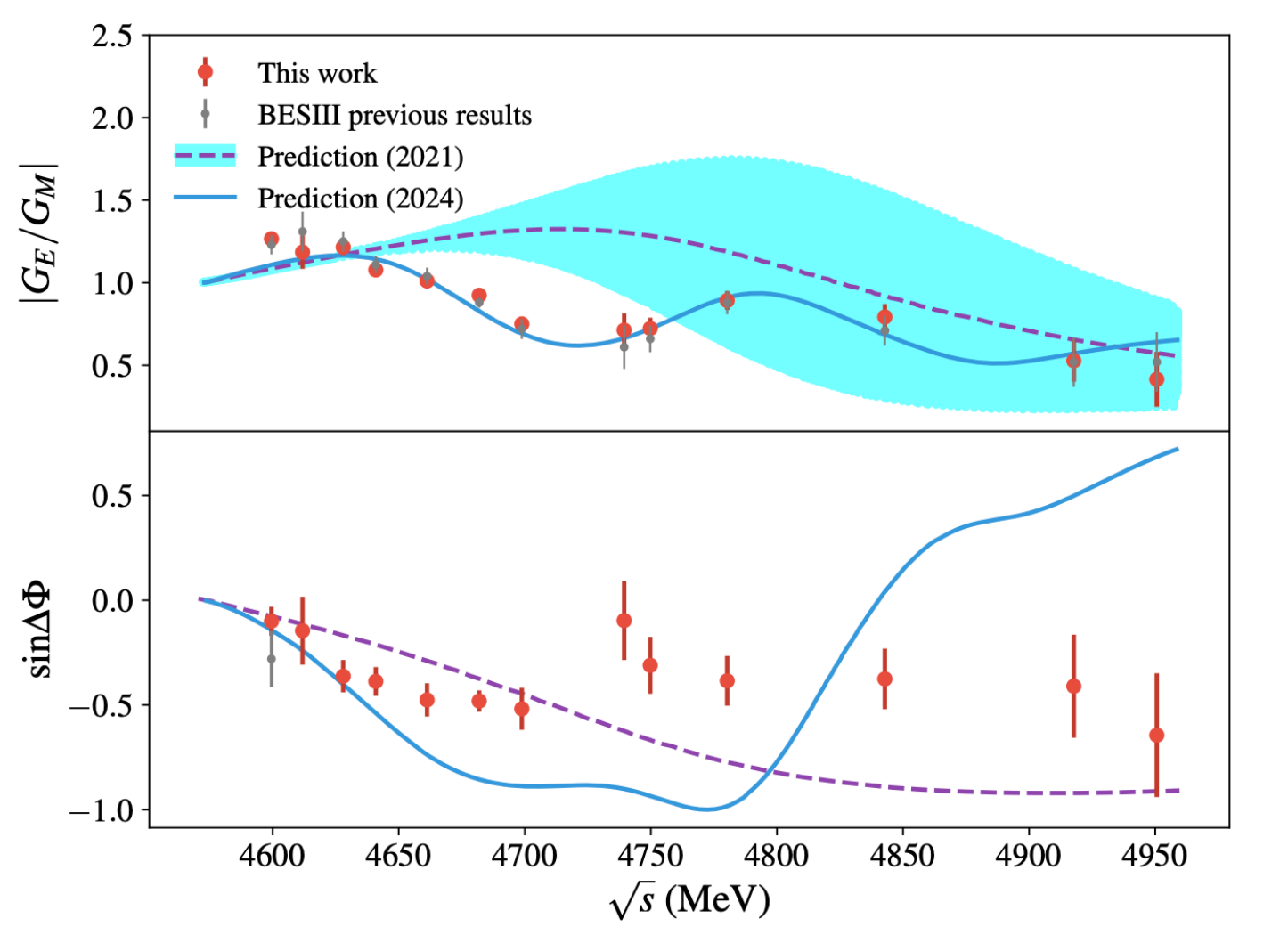}
    \end{tabular}

    \caption{(Left, Center) Fits of GNN output to determine $\Lambda_c^+\to n e^+\nu$ and $\Lambda_c^-\to \overline{n} e^-\nu$ signal yields from Ref.~\cite{BESIII:2024mgg}. (Right) Comparison of $|G_E|/|G_M|$ and $\sin\Delta\Phi$ determined in Ref.~\cite{BESIII:2025zbz} to predictions ~\cite{Wan:2021ncg,Chen:2023oqs}. }
    \label{fig:secondfig}
\end{figure}

The first observation of the semileptonic decay $\Lambda_c^+ \to ne^+\nu_e$ was achieved using a 4.5~fb$^{-1}$ sample collected at $E_{\rm CM} = 4.6$--$4.7$~GeV~\cite{BESIII:2024mgg}. This decay represents the first measurement of a $c \to d$ transition in the charmed baryon sector. The major experimental challenge is the background from $\Lambda_c^+ \to \Lambda[{\to} n\pi^0]e^+\nu_e$, where the $\pi^0$ is not reconstructed. To suppress this background, a graph neural network (GNN) was trained on control samples of $J/\psi$ decays to discriminate between neutrons and $\Lambda$ baryons decaying to $n\pi^0$. Signal yields were determined from a simultaneous fit to the corrected GNN output distributions for $\Lambda_c^+ \to ne^+\nu$ and the charge-conjugate mode $\bar{\Lambda}_c^- \to \bar{n}e^-\bar{\nu}$, shown in Fig.~\ref{fig:secondfig}. The measured branching fraction is $\mathcal{B}(\Lambda_c^+ \to ne^+\nu) = (0.357 \pm 0.034 \pm 0.014)\%$. Combining this result with lattice QCD predictions for the $\Lambda_c \to n$ differential decay rate~\cite{Meinel:2017ggx} and the $\Lambda_c^+$ lifetime from Belle~II~\cite{Belle-II:2022ggx} yields $|V_{cd}| = 0.208 \pm 0.011_{\rm exp} \pm 0.007_{\rm LQCD} \pm 0.001_{\tau_{\Lambda_c}}$, providing an independent determination of this CKM matrix element.

Electron-positron collisions allow for unique probes of baryonic structure through the study of polarization in $\ee\to H\overline H$, where $H$ is any baryon. The production cross section for a baryon pair, such as $e^+e^- \to \Lambda_c^+\bar{\Lambda}_c^-$, is parametrized by the electric and magnetic timelike form factors $G_E$ and $G_M$. If these form factors have a relative phase $\Delta\Phi$ and $|G_E|/|G_M| \neq 1$, the produced $\Lambda_c$ baryons can exhibit transverse polarization parameterized by
\begin{equation}
P_y(\theta_0) = \frac{3}{2(3+\alpha_0)}\sqrt{1-\alpha_0^2}\sin\theta_0\cos\theta_0\sin\Delta\Phi,
\end{equation}
where $\theta_0$ is the $\Lambda_c^+$ momentum angle in the $\ee$ rest grame and $\alpha_0 = \alpha_0(|G_E|/|G_M|)$ is a function of the form factor ratio. Using $6.4\invfb$ of data collected between $\Ecm=4.60-4.95\GeV$,  this polarization was measured in Ref.~\cite{BESIII:2025zbz} through the helicity angle distributions in $\Lambda_c^+$ decays to $pK^0_S$, $\Lambda\pi^+$, $\Sigma^0\pi^+$, $\Sigma^+\pi^-$, and $pK^-\pi^+$ (using decay parameters measured by the LHCb collaboration~\cite{LHCb:2023crj,LHCb:2022sck}). A simultaneous fit to the helicity angle distributions in all decay modes was performed at multiple center-of-mass energies to determine $|G_E|/|G_M|$ and $\sin\Delta\Phi$, shown in Fig.~\ref{fig:secondfig}. These measurements provide the first evidence for transverse polarization in charm baryon pair production. The results show good agreement with recent theoretical predictions~\cite{Wan:2021ncg,Chen:2023oqs} of the magnitude ratio $|G_E|/|G_M|$, but significant deviations are seen between the predictions and experimental determinations of $\sin\Delta\Phi$.

Additional recent measurements of $\Lambda_c^+$ decays at BESIII include the first observation of the singly-Cabibbo-suppressed decay $\Lambda_c^+ \to p\pi^0$~\cite{BESIII:2024cbr}, measurements of $\Lambda_c^+ \to \Lambda K^0_Sh^+$, $\Sigma K^0_Sh^+$, and $\Xi K^0_Sh^+$ ($h = \pi, K$) branching fractions~\cite{Hyperons}, the branching fractions for $\Lambda_c^+ \to \Sigma^+\eta$ and $\Lambda_c^+ \to \Sigma^+\eta'$~\cite{BESIII:2025vvd}, and an inclusive measurement of $\Lambda_c^+ \to K^0_SX$~\cite{KSIncl}.

In summary, BESIII has produced a wealth of precision measurements in charm semileptonic decays and charmed baryon physics using datasets collected at the $\psi(3770)$ resonance and at higher center-of-mass energies. These proceedings highlight the most precise single determination of $|V_{cs}|$ from $D \to K\ell\nu$ decays and stringent tests of lepton flavor universality in the charm sector; comprehensive studies of $D \to K\pi X\ell\nu$ transitions, the first measurement of the $D\to a_0(980)$ form factor; the first observation of the $c \to d$ transition in charmed baryons through $\Lambda_c^+ \to ne^+\nu$; and the first measurement of transverse polarization in $e^+e^- \to \Lambda_c^+\bar{\Lambda}_c^-$ production.

Much additional physics remains to be extracted from the existing BESIII datasets. Many new and updated measurements of $D^0$ and $D^+$ decays are underway with the full $20.3\invfb$ dataset collected at the \psipp resonance. Looking forward, the recently completed BEPCII upgrade will provide substantially higher instantaneous luminosity and extend the accessible center-of-mass energy range to 5.6~GeV, enabling studies of $\Sigma_c$, $\Xi_c$, and $\Omega_c$ baryon pair production. An additional 9~fb$^{-1}$ of $\Lambda_c^+$ data is planned for 2025--2026, with data collection above 5~GeV scheduled for 2028. These future datasets will enable further precision tests of the Standard Model and searches for new physics in the charm sector.

% \begin{figure}[h]
% \centering
% \includegraphics[width=0.45\textwidth]{figures/detector.pdf}
% \caption{Cross-sectional view of the BESIII detector.}
% \label{fig:besiii}
% \end{figure}

% \section{Recent measurements of $\ee\to\psi(3770)\to \DD$}

\bibliographystyle{LHCb}
\bibliography{main}

@PREAMBLE{
 "\providecommand{\noopsort}[1]{}" 
 # "\providecommand{\singleletter}[1]{#1}%" 
}

@article{BESIIIDetector,
    author = "Ablikim, M. and others",
    collaboration = "BESIII Collaboration",
    title = "{Design and Construction of the BESIII Detector}",
    eprint = "0911.4960",
    archivePrefix = "arXiv",
    primaryClass = "physics.ins-det",
    doi = "10.1016/j.nima.2009.12.050",
    journal = "Nucl. Instrum. Meth. A",
    volume = "614",
    pages = "345--399",
    year = "2010"
}

@inproceedings{BEPCII,
    author = "Yu, Chenghui and others",
    title = "{BEPCII Performance and Beam Dynamics Studies on Luminosity}",
    booktitle = "{7th International Particle Accelerator Conference}",
    doi = "10.18429/JACoW-IPAC2016-TUYA01",
    pages = "TUYA01",
    year = "2016"
}

@article{ParticleDataGroup:2024cfk,
    author = "Navas, S. and others",
    collaboration = "Particle Data Group",
    title = "{Review of particle physics}",
    doi = "10.1103/PhysRevD.110.030001",
    journal = "Phys. Rev. D",
    volume = "110",
    number = "3",
    pages = "030001",
    year = "2024"
}

@article{BESIII:2020nme,
    author = "Ablikim, M. and others",
    collaboration = "BESIII Collaboration",
    title = "{Future Physics Programme of BESIII}",
    eprint = "1912.05983",
    archivePrefix = "arXiv",
    primaryClass = "hep-ex",
    reportNumber = "HEP-Physics-Report-BESIII-2019-12-13",
    doi = "10.1088/1674-1137/44/4/040001",
    journal = "Chin. Phys. C",
    volume = "44",
    number = "4",
    pages = "040001",
    year = "2020"
}

@article{gb8v-4rnh,
  title = {Precision Measurement of the Branching Fraction of ${D}^{+}\rightarrow{\mu}^{+}{\nu}_{\mu}$},
  author = {Ablikim, M. and Achasov, M. N. and Adlarson, P. and Afedulidis, O. and Ai, X. C. and Aliberti, R. and Amoroso, A. and An, Q. and Bai, Y. and Bakina, O. and Balossino, I. and Ban, Y. and Bao, H.-R. and Batozskaya, V. and Begzsuren, K. and Berger, N. and Berlowski, M. and Bertani, M. and Bettoni, D. and Bianchi, F. and Bianco, E. and Bortone, A. and Boyko, I. and Briere, R. A. and Brueggemann, A. and Cai, H. and Cai, X. and Calcaterra, A. and Cao, G. F. and Cao, N. and Cetin, S. A. and Chang, J. F. and Che, G. R. and Chelkov, G. and Chen, C. and Chen, C. H. and Chen, Chao and Chen, G. and Chen, H. S. and Chen, H. Y. and Chen, M. L. and Chen, S. J. and Chen, S. L. and Chen, S. M. and Chen, T. and Chen, X. R. and Chen, X. T. and Chen, Y. B. and Chen, Y. Q. and Chen, Z. J. and Chen, Z. Y. and Choi, S. K. and Cibinetto, G. and Cossio, F. and Cui, J. J. and Dai, H. L. and Dai, J. P. and Dbeyssi, A. and de Boer, R. E. and Dedovich, D. and Deng, C. Q. and Deng, Z. Y. and Denig, A. and Denysenko, I. and Destefanis, M. and De Mori, F. and Ding, B. and Ding, X. X. and Ding, Y. and Ding, Y. and Dong, J. and Dong, L. Y. and Dong, M. Y. and Dong, X. and Du, M. C. and Du, S. X. and Duan, Y. Y. and Duan, Z. H. and Egorov, P. and Fan, Y. H. and Fang, J. and Fang, J. and Fang, S. S. and Fang, W. X. and Fang, Y. and Fang, Y. Q. and Farinelli, R. and Fava, L. and Feldbauer, F. and Felici, G. and Feng, C. Q. and Feng, J. H. and Feng, Y. T. and Fritsch, M. and Fu, C. D. and Fu, J. L. and Fu, Y. W. and Gao, H. and Gao, X. B. and Gao, Y. N. and Gao, Yang and Garbolino, S. and Garzia, I. and Ge, L. and Ge, P. T. and Ge, Z. W. and Geng, C. and Gersabeck, E. M. and Gilman, A. and Goetzen, K. and Gong, L. and Gong, W. X. and Gradl, W. and Gramigna, S. and Greco, M. and Gu, M. H. and Gu, Y. T. and Guan, C. Y. and Guo, A. Q. and Guo, L. B. and Guo, M. J. and Guo, R. P. and Guo, Y. P. and Guskov, A. and Gutierrez, J. and Han, K. L. and Han, T. T. and Hanisch, F. and Hao, X. Q. and Harris, F. A. and He, K. K. and He, K. L. and Heinsius, F. H. and Heinz, C. H. and Heng, Y. K. and Herold, C. and Holtmann, T. and Hong, P. C. and Hou, G. Y. and Hou, X. T. and Hou, Y. R. and Hou, Z. L. and Hu, B. Y. and Hu, H. M. and Hu, J. F. and Hu, S. L. and Hu, T. and Hu, Y. and Huang, G. S. and Huang, K. X. and Huang, L. Q. and Huang, X. T. and Huang, Y. P. and Huang, Y. S. and Hussain, T. and H\"olzken, F. and H\"usken, N. and in der Wiesche, N. and Jackson, J. and Janchiv, S. and Jeong, J. H. and Ji, Q. and Ji, Q. P. and Ji, W. and Ji, X. B. and Ji, X. L. and Ji, Y. Y. and Jia, X. Q. and Jia, Z. K. and Jiang, D. and Jiang, H. B. and Jiang, P. C. and Jiang, S. S. and Jiang, T. J. and Jiang, X. S. and Jiang, Y. and Jiao, J. B. and Jiao, J. K. and Jiao, Z. and Jin, S. and Jin, Y. and Jing, M. Q. and Jing, X. M. and Johansson, T. and Kabana, S. and Kalantar-Nayestanaki, N. and Kang, X. L. and Kang, X. S. and Kavatsyuk, M. and Ke, B. C. and Khachatryan, V. and Khoukaz, A. and Kiuchi, R. and Kolcu, O. B. and Kopf, B. and Kuessner, M. and Kui, X. and Kumar, N. and Kupsc, A. and K\"uhn, W. and Lane, J. J. and Lavezzi, L. and Lei, T. T. and Lei, Z. H. and Lellmann, M. and Lenz, T. and Li, C. and Li, C. and Li, C. H. and Li, Cheng and Li, D. M. and Li, F. and Li, G. and Li, H. B. and Li, H. J. and Li, H. N. and Li, Hui and Li, J. R. and Li, J. S. and Li, K. and Li, K. L. and Li, L. J. and Li, L. K. and Li, Lei and Li, M. H. and Li, P. R. and Li, Q. M. and Li, Q. X. and Li, R. and Li, S. X. and Li, T. and Li, W. D. and Li, W. G. and Li, X. and Li, X. H. and Li, X. L. and Li, X. Y. and Li, X. Z. and Li, Y. G. and Li, Z. J. and Li, Z. Y. and Liang, C. and Liang, H. and Liang, H. and Liang, Y. F. and Liang, Y. T. and Liao, G. R. and Liao, Y. P. and Libby, J. and Limphirat, A. and Lin, C. C. and Lin, D. X. and Lin, T. and Liu, B. J. and Liu, B. X. and Liu, C. and Liu, C. X. and Liu, F. and Liu, F. H. and Liu, Feng and Liu, G. M. and Liu, H. and Liu, H. B. and Liu, H. H. and Liu, H. M. and Liu, Huihui and Liu, J. B. and Liu, J. Y. and Liu, K. and Liu, K. Y. and Liu, Ke and Liu, L. and Liu, L. C. and Liu, Lu and Liu, M. H. and Liu, P. L. and Liu, Q. and Liu, S. B. and Liu, T. and Liu, W. K. and Liu, W. M. and Liu, X. and Liu, X. and Liu, Y. and Liu, Y. and Liu, Y. B. and Liu, Z. A. and Liu, Z. D. and Liu, Z. Q. and Lou, X. C. and Lu, F. X. and Lu, H. J. and Lu, J. G. and Lu, X. L. and Lu, Y. and Lu, Y. P. and Lu, Z. H. and Luo, C. L. and Luo, J. R. and Luo, M. X. and Luo, T. and Luo, X. L. and Lyu, X. R. and Lyu, Y. F. and Ma, F. C. and Ma, H. and Ma, H. L. and Ma, J. L. and Ma, L. L. and Ma, L. R. and Ma, M. M. and Ma, Q. M. and Ma, R. Q. and Ma, T. and Ma, X. T. and Ma, X. Y. and Ma, Y. and Ma, Y. M. and Maas, F. E. and Maggiora, M. and Malde, S. and Mao, Y. J. and Mao, Z. P. and Marcello, S. and Meng, Z. X. and Messchendorp, J. G. and Mezzadri, G. and Miao, H. and Min, T. J. and Mitchell, R. E. and Mo, X. H. and Moses, B. and Muchnoi, N. Yu. and Muskalla, J. and Nefedov, Y. and Nerling, F. and Nie, L. S. and Nikolaev, I. B. and Ning, Z. and Nisar, S. and Niu, Q. L. and Niu, W. D. and Niu, Y. and Olsen, S. L. and Ouyang, Q. and Pacetti, S. and Pan, X. and Pan, Y. and Pathak, A. and Pei, Y. P. and Pelizaeus, M. and Peng, H. P. and Peng, Y. Y. and Peters, K. and Ping, J. L. and Ping, R. G. and Plura, S. and Prasad, V. and Qi, F. Z. and Qi, H. and Qi, H. R. and Qi, M. and Qi, T. Y. and Qian, S. and Qian, W. B. and Qiao, C. F. and Qiao, X. K. and Qin, J. J. and Qin, L. Q. and Qin, L. Y. and Qin, X. P. and Qin, X. S. and Qin, Z. H. and Qiu, J. F. and Qu, Z. H. and Redmer, C. F. and Ren, K. J. and Rivetti, A. and Rolo, M. and Rong, G. and Rosner, Ch. and Ruan, S. N. and Salone, N. and Sarantsev, A. and Schelhaas, Y. and Schoenning, K. and Scodeggio, M. and Shan, K. Y. and Shan, W. and Shan, X. Y. and Shang, Z. J. and Shangguan, J. F. and Shao, L. G. and Shao, M. and Shen, C. P. and Shen, H. F. and Shen, W. H. and Shen, X. Y. and Shi, B. A. and Shi, H. and Shi, H. C. and Shi, J. L. and Shi, J. Y. and Shi, Q. Q. and Shi, S. Y. and Shi, X. and Song, J. J. and Song, T. Z. and Song, W. M. and Song, Y. J. and Song, Y. X. and Sosio, S. and Spataro, S. and Stieler, F. and Su, S. S. and Su, Y. J. and Sun, G. B. and Sun, G. X. and Sun, H. and Sun, H. K. and Sun, J. F. and Sun, K. and Sun, L. and Sun, S. S. and Sun, T. and Sun, W. Y. and Sun, Y. and Sun, Y. J. and Sun, Y. Z. and Sun, Z. Q. and Sun, Z. T. and Tang, C. J. and Tang, G. Y. and Tang, J. and Tang, M. and Tang, Y. A. and Tao, L. Y. and Tao, Q. T. and Tat, M. and Teng, J. X. and Thoren, V. and Tian, W. H. and Tian, Y. and Tian, Z. F. and Uman, I. and Wan, Y. and Wang, S. J. and Wang, B. and Wang, B. L. and Wang, Bo and Wang, D. Y. and Wang, F. and Wang, H. J. and Wang, J. J. and Wang, J. P. and Wang, K. and Wang, L. L. and Wang, M. and Wang, N. Y. and Wang, S. and Wang, S. and Wang, T. and Wang, T. J. and Wang, W. and Wang, W. and Wang, W. P. and Wang, X. and Wang, X. F. and Wang, X. J. and Wang, X. L. and Wang, X. N. and Wang, Y. and Wang, Y. D. and Wang, Y. F. and Wang, Y. L. and Wang, Y. N. and Wang, Y. Q. and Wang, Yaqian and Wang, Yi and Wang, Z. and Wang, Z. L. and Wang, Z. Y. and Wang, Ziyi and Wei, D. H. and Weidner, F. and Wen, S. P. and Wen, Y. R. and Wiedner, U. and Wilkinson, G. and Wolke, M. and Wollenberg, L. and Wu, C. and Wu, J. F. and Wu, L. H. and Wu, L. J. and Wu, X. and Wu, X. H. and Wu, Y. and Wu, Y. H. and Wu, Y. J. and Wu, Z. and Xia, L. and Xian, X. M. and Xiang, B. H. and Xiang, T. and Xiao, D. and Xiao, G. Y. and Xiao, S. Y. and Xiao, Y. L. and Xiao, Z. J. and Xie, C. and Xie, X. H. and Xie, Y. and Xie, Y. G. and Xie, Y. H. and Xie, Z. P. and Xing, T. Y. and Xu, C. F. and Xu, C. J. and Xu, G. F. and Xu, H. Y. and Xu, M. and Xu, Q. J. and Xu, Q. N. and Xu, W. and Xu, W. L. and Xu, X. P. and Xu, Y. and Xu, Y. C. and Xu, Z. S. and Yan, F. and Yan, L. and Yan, W. B. and Yan, W. C. and Yan, X. Q. and Yang, H. J. and Yang, H. L. and Yang, H. X. and Yang, T. and Yang, Y. and Yang, Y. F. and Yang, Y. F. and Yang, Y. X. and Yang, Z. W. and Yao, Z. P. and Ye, M. and Ye, M. H. and Yin, J. H. and Yin, Junhao and You, Z. Y. and Yu, B. X. and Yu, C. X. and Yu, G. and Yu, J. S. and Yu, M. C. and Yu, T. and Yu, X. D. and Yu, Y. C. and Yuan, C. Z. and Yuan, J. and Yuan, J. and Yuan, L. and Yuan, S. C. and Yuan, Y. and Yuan, Z. Y. and Yue, C. X. and Zafar, A. A. and Zeng, F. R. and Zeng, S. H. and Zeng, X. and Zeng, Y. and Zeng, Y. J. and Zeng, Y. J. and Zhai, X. Y. and Zhai, Y. C. and Zhan, Y. H. and Zhang, A. Q. and Zhang, B. L. and Zhang, B. X. and Zhang, D. H. and Zhang, G. Y. and Zhang, H. and Zhang, H. and Zhang, H. C. and Zhang, H. H. and Zhang, H. H. and Zhang, H. Q. and Zhang, H. R. and Zhang, H. Y. and Zhang, J. and Zhang, J. and Zhang, J. J. and Zhang, J. L. and Zhang, J. Q. and Zhang, J. S. and Zhang, J. W. and Zhang, J. X. and Zhang, J. Y. and Zhang, J. Z. and Zhang, Jianyu and Zhang, L. M. and Zhang, Lei and Zhang, P. and Zhang, Q. Y. and Zhang, R. Y. and Zhang, S. H. and Zhang, Shulei and Zhang, X. D. and Zhang, X. M. and Zhang, X. Y. and Zhang, X. Y. and Zhang, Y. and Zhang, Y. and Zhang, Y. T. and Zhang, Y. H. and Zhang, Y. M. and Zhang, Yan and Zhang, Z. D. and Zhang, Z. H. and Zhang, Z. L. and Zhang, Z. Y. and Zhang, Z. Y. and Zhang, Z. Z. and Zhao, G. and Zhao, J. Y. and Zhao, J. Z. and Zhao, L. and Zhao, Lei and Zhao, M. G. and Zhao, N. and Zhao, R. P. and Zhao, S. J. and Zhao, Y. B. and Zhao, Y. X. and Zhao, Z. G. and Zhemchugov, A. and Zheng, B. and Zheng, B. M. and Zheng, J. P. and Zheng, W. J. and Zheng, Y. H. and Zhong, B. and Zhong, X. and Zhou, H. and Zhou, J. Y. and Zhou, L. P. and Zhou, S. and Zhou, X. and Zhou, X. K. and Zhou, X. R. and Zhou, X. Y. and Zhou, Y. Z. and Zhou, Z. C. and Zhu, A. N. and Zhu, J. and Zhu, K. and Zhu, K. J. and Zhu, K. S. and Zhu, L. and Zhu, L. X. and Zhu, S. H. and Zhu, T. J. and Zhu, W. D. and Zhu, Y. C. and Zhu, Z. A. and Zou, J. H. and Zu, J.},
  collaboration = {BESIII Collaboration},
  journal = {Phys. Rev. Lett.},
  volume = {135},
  issue = {6},
  pages = {061801},
  numpages = {10},
  year = {2025},
  month = {Aug},
  publisher = {American Physical Society},
  doi = {10.1103/gb8v-4rnh},
  url = {https://link.aps.org/doi/10.1103/gb8v-4rnh}
}

@article{BESIII:2024vlt,
    author = "Ablikim, Medina and others",
    collaboration = "BESIII Collaboration",
    title = "{Measurement of the branching fraction of $D^{+} \rightarrow \tau^{+}\nu_{\tau}$}",
    eprint = "2410.20063",
    archivePrefix = "arXiv",
    primaryClass = "hep-ex",
    doi = "10.1007/JHEP01(2025)089",
    journal = "JHEP",
    volume = "01",
    pages = "089",
    year = "2025"
}

@article{ParticleDataGroup:2020ssz,
    author = "Zyla, P. A. and others",
    collaboration = "Particle Data Group",
    title = "{Review of Particle Physics}",
    doi = "10.1093/ptep/ptaa104",
    journal = "PTEP",
    volume = "2020",
    number = "8",
    pages = "083C01",
    year = "2020"
}

@article{Riggio:2017zwh,
    author = "Riggio, L. and Salerno, G. and Simula, S.",
    title = "{Extraction of $|V_{cd}|$ and $|V_{cs}|$ from experimental decay rates using lattice QCD $D \rightarrow \pi(K) \ell \nu$ form factors}",
    eprint = "1706.03657",
    archivePrefix = "arXiv",
    primaryClass = "hep-lat",
    reportNumber = "PREPRINT-RM3-TH-17-7, preprint RM3-TH/17-7",
    doi = "10.1140/epjc/s10052-018-5943-5",
    journal = "Eur. Phys. J. C",
    volume = "78",
    number = "6",
    pages = "501",
    year = "2018"
}

@article{MARK-III:1987jsm,
    author = "Adler, J. and others",
    editor = "Tran Thanh Van, J.",
    collaboration = "MARK-III Collaboration",
    title = "{A Reanalysis of Charmed d Meson Branching Fractions}",
    reportNumber = "SLAC-PUB-4291",
    journal = {Phys. Rev. Lett.},
    doi = "10.1103/PhysRevLett.60.89",
    pages = "153--160",
    year = "1987"
}

@article{ARGUS:1990hfq,
    author = "Albrecht, H. and others",
    collaboration = "ARGUS Collaboration",
    title = "{Search for Hadronic $b \rightarrow u$ Decays}",
    reportNumber = "DESY-90-008",
    doi = "10.1016/0370-2693(90)91293-K",
    journal = "Phys. Lett. B",
    volume = "241",
    pages = "278--282",
    year = "1990"
}

@article{BESIII:2024slx,
    author = "Ablikim, Medina and others",
    collaboration = "BESIII Collaboration",
    title = "{Improved measurements of $D^0 \rightarrow K^-\ell^+\nu_{\ell}$ and $D^+ \rightarrow \bar{K}^0\ell^+\nu_{\ell}$}",
    eprint = "2408.09087",
    archivePrefix = "arXiv",
    primaryClass = "hep-ex",
    doi = "10.1103/PhysRevD.110.112006",
    journal = "Phys. Rev. D",
    volume = "110",
    number = "11",
    pages = "112006",
    year = "2024"
}

@article{Parrott:2022rgu,
    author = "Parrott, W. G. and Bouchard, C. and Davies, C. T. H.",
    collaboration = "HPQCD collaboration, HPQCD",
    title = "{$B \rightarrow K$ and $D \rightarrow K$ form factors from fully relativistic lattice QCD}",
    eprint = "2207.12468",
    archivePrefix = "arXiv",
    primaryClass = "hep-lat",
    doi = "10.1103/PhysRevD.107.014510",
    journal = "Phys. Rev. D",
    volume = "107",
    number = "1",
    pages = "014510",
    year = "2023"
}

@article{FermilabLattice:2022gku,
    author = "Bazavov, Alexei and others",
    collaboration = "Fermilab Lattice, MILC Collaboration",
    title = "{D-meson semileptonic decays to pseudoscalars from four-flavor lattice QCD}",
    eprint = "2212.12648",
    archivePrefix = "arXiv",
    primaryClass = "hep-lat",
    reportNumber = "MIT-CTP/5513, FERMILAB-PUB-22-943-T",
    doi = "10.1103/PhysRevD.107.094516",
    journal = "Phys. Rev. D",
    volume = "107",
    number = "9",
    pages = "094516",
    year = "2023"
}

@article{BESIII:2024xjf,
    author = "Ablikim, Medina and others",
    collaboration = "BESIII Collaboration",
    title = "{Study of the semileptonic decay $D^0\rightarrow \bar{K}^0\pi^{-}e^{+}\nu_e$}",
    eprint = "2412.10803",
    archivePrefix = "arXiv",
    primaryClass = "hep-ex",
    doi = "10.1007/JHEP03(2025)197",
    journal = "JHEP",
    volume = "03",
    pages = "197",
    year = "2025"
}

@article{BESIII:2025yot,
    author = "Ablikim, Medina and others",
    collaboration = "BESIII Collaboration",
    title = "{Observation of $D^+\rightarrow \bar{K}_1(1270)^0\mu^+\nu_\mu$ and $D^0\rightarrow K_1(1270)^-\mu^+\nu_\mu$}",
    eprint = "2502.03828",
    archivePrefix = "arXiv",
    primaryClass = "hep-ex",
    doi = "10.1103/PhysRevD.111.L071101",
    month = "2",
    year = "2025"
}

@article{BESIII:2025ujq,
    author = "Ablikim, M. and others",
    collaboration = "BESIII Collaboration",
    title = "{Study of the light scalar $a_0(980)$ through the decay $D^0 \rightarrow a_0(980)^-e^+ \nu_e$ with $a_0(980)^- \rightarrow \eta\pi^-$}",
    doi = "10.1103/PhysRevD.111.L091501",
    journal = "Phys. Rev. D",
    volume = "111",
    number = "9",
    pages = "L091501",
    year = "2025"
}

@article{Soni:2020sgn,
    author = "Soni, Nakul R. and Gadaria, Akshay N. and Patel, Janaki J. and Pandya, Jignesh N.",
    title = "{Semileptonic Decays of Charmed Mesons to Light Scalar Mesons}",
    eprint = "2001.10195",
    archivePrefix = "arXiv",
    primaryClass = "hep-ph",
    doi = "10.1103/PhysRevD.102.016013",
    journal = "Phys. Rev. D",
    volume = "102",
    number = "1",
    pages = "016013",
    year = "2020"
}

@article{Momeni:2022gqb,
    author = "Momeni, S. and Saghebfar, M.",
    title = "{Semileptonic $D$ meson decays to the vector, axial vector and scalar mesons in Hard-Wall AdS/QCD correspondence}",
    doi = "10.1140/epjc/s10052-022-10413-x",
    journal = "Eur. Phys. J. C",
    volume = "82",
    number = "5",
    pages = "473",
    year = "2022"
}

@article{Cheng:2017fkw,
    author = "Cheng, Xiao-Dong and Li, Hai-Bo and Wei, Bin and Xu, Yu-Guo and Yang, Mao-Zhi",
    title = "{Study of $D \rightarrow a_0 (980) e^+ \nu_e$ decay in the light-cone sum rules approach}",
    eprint = "1706.01019",
    archivePrefix = "arXiv",
    primaryClass = "hep-ph",
    doi = "10.1103/PhysRevD.96.033002",
    journal = "Phys. Rev. D",
    volume = "96",
    number = "3",
    pages = "033002",
    year = "2017"
}

@article{Huang:2021owr,
    author = "Huang, Qi and Sun, Yan-Jun and Gao, Di and Zhao, Guo-Hua and Wang, Bin and Hong, Wei",
    title = "{Study of form factors and branching ratios for $D\rightarrow S,Al\bar{\nu_{l}}$ with light-cone sum rules}",
    eprint = "2102.12241",
    archivePrefix = "arXiv",
    primaryClass = "hep-ph",
    month = "2",
    year = "2021"
}

\end{document}